\documentclass[11pt,a4paper]{article}
\pdfoutput=1

\usepackage{amsmath}
\usepackage{amsfonts}
\usepackage{amsthm}
\usepackage{amssymb}
\usepackage{amscd}
\usepackage[british]{babel}
\usepackage{graphicx}
\usepackage{psfrag}
\usepackage{epsfig}
\usepackage{rotating}
\usepackage{times}

\theoremstyle{plain}

\newtheorem{remark}{Remark}[section]

\newcommand{\boxend}{\flushright{$\Box$}}

\newcommand{\N}{{\mathbb N}}               
\newcommand{\Z}{{\mathbb Z}}               
\newcommand{\R}{{\mathbb R}}               

\begin{document}

\title{Does loop quantum cosmology replace the big rip singularity by a non-singular bounce?}

\author{Jaume de Haro$^{a,}$\footnote{E-mail: jaime.haro@upc.edu}
}

\maketitle

{$^a$Departament de Matem\`atica Aplicada I, Universitat
Polit\`ecnica de Catalunya, Diagonal 647, 08028 Barcelona, Spain \\
}

\thispagestyle{empty}

\begin{abstract} It is stated that holonomy corrections in
loop quantum cosmology introduce a modification  in  Friedmann's equation which prevent
 the big rip singularity. Recently in \cite{h12} it has been proved that  this modified Friedmann equation is
obtained in an inconsistent way, what
means that the  results deduced  from it, in particular the big rip singularity avoidance, are not justified. The problem is
that holonomy corrections modify the gravitational part of the Hamiltonian of the system  leading, after Legendre's transformation,  to
a non covariant Lagrangian which is in contradiction with one of the main principles of General Relativity. A more consistent way to deal with the big
rip singularity avoidance is
to disregard modification in the gravitational part of the Hamiltonian, and
only  consider inverse volume effects \cite{bo02a}. In this case we will see that, not like  the big bang singularity,  the
big rip singularity survives in loop quantum cosmology.
Another way to deal with the big rip avoidance is to take into account geometric quantum effects
given by the the Wheeler-De Witt equation. In that case, even though the wave packets spread, the
expectation values satisfy the same equations as their classical analogues. Then, following the viewpoint adopted in loop quantum cosmology, one
can conclude that the big rip singularity survives when one takes into account these quantum effects.
However, the spreading of the wave packets prevents the recover of the semiclassical time, and thus, one might conclude that
the classical
evolution of the universe come to and end before the big rip is reached. This is not conclusive  because. as we will see, it always exists other external times
that allows us to define  the  classical and
quantum  evolution of the universe up to the big rip singularity.
\end{abstract}

\vspace{0.5cm}

{\bf Pacs numbers:} 98.80.Qc, 04.20.Dw, 04.60.Pp, 04.20.Fy

\vspace{0.5cm}

\noindent{\it 1.- Introduction}---
Observations from  the spacecraft {\it Wilkinson Microwave Anisotropy Probe} (WMAP) indicate
 that our universe
would be dominated by ``phantom energy``. This fact would be explained assuming the presence of dark energy with a negative pressure, which
could be modeled by a barotropic fluid with equation of state (EoS)  $P=\omega \rho$, with $\omega=-1.10\pm 0.14$ (see \cite{k11}).
However, a universe containing
dark energy  modeled by  the equation $P=\omega \rho$  with $\omega<-1$ leads  to a classical future singularity called {\it big rip}
\cite{s00,ckw03}.

This singularity is at classical level, i.e., it appears solving the classical Raychaudury equation. However,  if one take into
account quantum effects it could be avoided, that is what happens in loop quantum cosmology where
a good number of papers have
shown that this singularity is replaced by a non-singular bounce (see for instance \cite{svv06,sst06,nw07,sg07,s09,lw10}).

The main idea is that loop quantum cosmology assumes a discrete nature of space which leads, at quantum level, to consider a Hilbert
space  where
quantum states are represented by
 almost periodic functions of the dynamical part of the connection.
Unfortunately, the connection variable doesn't correspond to a well defined quantum operator in this Hilbert space, what leads to
 re-express the gravitational part of the
Hamiltonian in terms
of almost periodic function,  which could be done from a  process of regularization.
This new regularized Hamiltonian
 introduces a quadratic modification ($\rho^2$)
in the Friedmann equation
at high energies which give rise to a bounce when the energy density becomes equal to a critical value of the order of the Planck energy density.

Recently, we have shown in \cite{h12}
that the regularized Hamiltonian, after Legendre's transformation, gives rise to a non covariant Lagrangian, what contradicts the main principles of
General Relativity, in other words, the regularized Hamiltonian comes from a non covariant Lagrangian, what means that it is un-physical.
As a consequence,
the
modified Friedmann equation, which is obtained from this regularized Hamiltonian, is incorrect. And thus,  the current statement that,
in loop quantum cosmology,
 the big rip singularity is
replaced by a non-singular bounce, could be wrong because it is deduced from an incorrect modified Friedmann equation.

To sort out this situation, it seems clair that one cannot change the gravitational part of the Hamiltonian (or it has to be changed by
another one that doesn't lead to a non covariant Lagrangian), what gives rise to disregard the
Hilbert space of almost periodic functions. In this way we review  Bojowald's early work where the author only takes into account
inverse volume correction in the energy density. These corrections lead to a new modified Friedmann equation where the big bang singularity
is avoided, but the big rip one survives.

The big rip singularity could also be studied in the framework of the Wheeler-DeWitt equation. In this case, since one assumes a continuous nature of the space, one does not need to change the gravitational part of the Hamiltonian, and thus, one does not have any contradiction with the principles of General Relativity.
 The Wheeler-DeWitt equation is obtained from the quantization of the so-called Hamiltonian constrain, leading to an equation where the cosmic 
time does not appear (this is the problem of the time in quantum gravity). Sometimes it is claimed that  the  
validity of the semiclassical approximation is needed   to recover the classical time
(see for example \cite{dks06,h09,v89,h87,lr79}), but in loop quantum cosmology  another
viewpoint is adopted \cite{aps06}: one can transform the Wheeler-DeWitt equation in a Schr\"odinger one taking the "square root" of the Wheeler-DeWitt equation, as a consequence, one has chose a classical
external time, that could be used to describe both, the classical and quantum dynamics of the universe.   Using this Schr\"odinger equation one can show that the expectation values satisfy
the same dynamical equations that their classical analogues, which in particular means that the big rip singularity is not avoided in the framework of the Wheeler-DeWitt equation.

The paper is organized as follows: In Section II we review  classical cosmology and the big rip singularity for a barotropic fluid with
EoS $P=\omega\rho$ and  $\omega<-1$. In Section III we give, from a mathematical point of view, a detailed discussion of the quantization in
loop cosmology. Section IV is devoted to a critical study of the effective formulation of loop quantum cosmology. Finally, in Section V we consider a toy
model of phantom cosmology, which allows us to show that neither
the Wheeler-DeWitt equation nor  loop quantum
cosmology avoid the big rip singularity.

The units used in the paper are
$\hbar=c=M_p=1$ being $M_p=\left(\frac{\hbar c}{8\pi G}\right)^{1/2}$ the reduced Planck mass.

\vspace{0.5cm}

\noindent{\it 2.- Classical cosmology and the big rip singularity}---
In this Section we consider
 the flat FRW space-time  filled by a perfect fluid with the EoS,  $P=\omega \rho$ (being $\omega<-1$).
In cosmology, the classical equations  are obtained from the Lagrangian
${\mathcal L}=\frac{1}{2}Ra^3-\rho a^3$  where $R=6(\dot{H}+2H^2)$ is the scalar curvature, $a$ is the
scale factor and $H=\frac{\dot{a}}{a}$ is the Hubble parameter.

This Lagrangian can be written as follows ${\mathcal L}=3\left(\frac{d{(\dot{a} a^2)}}{dt}-\dot{a}^2 a\right)-\rho a^3$, what means that the same theory
is obtained avoiding the total derivative, which gives the Lagrangian ${\mathcal L}_{E}=-3H^2 a^3-\rho a^3$. The conjugate momentum is  then given by
$p=\frac{\partial {\mathcal L}_{E}}{\partial\dot{a}}=-6Ha^2$,
 and thus the Hamiltonian is
\begin{eqnarray}\label{1}
{\mathcal H}_{E}=
\dot{a}p- {\mathcal L}_{E}=-3H^2a^3+\rho a^3.\end{eqnarray}

It's well known that in general relativity   the Hamiltonian is constrained to be zero,
which gives the classical
Friedmann equation:
\begin{eqnarray}\label{a2}
H^2=\rho/3.
\end{eqnarray}

This equation together with the conservation equation
\begin{eqnarray}\label{2}
 d(\rho a^3)=-Pd(a^3)\Longleftrightarrow
\dot{\rho}=-3H(1+\omega)\rho,
\end{eqnarray}
are the dynamical equations that describe the classical evolution of our universe.

Taking the derivative of the Friedmann equation with respect to the time and using the conservation equation one gets
the so-called Raychaudury equation
\begin{eqnarray}
 \dot{H}=-\frac{3}{2}(1+\omega)H^2,
\end{eqnarray}
which solution is given by:
\begin{eqnarray}\label{a1}
{H}(t)=\frac{2}{3(1+\omega)}\frac{1}{t-t_{rip}},
\end{eqnarray}
where $t_{rip}\equiv t_0-\frac{2}{3H_0(1+\omega)}$ is the time at which the big rip singularity is reached, and $H_0=H(t_0)$ is
the current value of the Hubble parameter.

From the classical Friedmann equation one also obtain
\begin{eqnarray}
 \rho(t)=\frac{4}{3(1+\omega)^2}\frac{1}{(t-t_{rip})^2}.
\end{eqnarray}

Since observations from WMAP give the approximate values $H_0\cong 2.3\times 10^{-18}\frac{1}{second}$ and $\omega\cong-1,1$,  one can conclude that the time
from the present to the big rip is approximately $90$ billion years.

\vspace{0.5cm}

\noindent{\it 3.- Loop quantum cosmology}---
The old quantization of loop quantum cosmology
was done using two canonically conjugate variables, one of them was
the dynamical part of the connection, namely ${\mathfrak c}$, and the other one was
the dynamical part of the triad, namely ${\mathfrak p}$. These variables are related with the scalar factor and
the extrinsic curvature $K=\frac{1}{2}\dot{a}$ by the relations
(see for instance \cite{abl03})
\begin{eqnarray}
 |{\mathfrak p}|=a^2, \qquad {\mathfrak c}=2\gamma K={\gamma}\dot{a},
\end{eqnarray}
where $\gamma$ is the Barbero-Immirzi parameter,
and their Poisson bracket is given by $\{{\mathfrak c},{\mathfrak p}\}=\frac{\partial {\mathfrak c}}{\partial a}\frac{\partial {\mathfrak p}}{\partial p}-
\frac{\partial {\mathfrak c}}{\partial p}\frac{\partial {\mathfrak p}}{\partial a}=\frac{\gamma}{3}\mbox{sgn}({\mathfrak p})$.

Alternatively, one can introduce the new canonically conjugate variables \cite{as11,s09a}:
\begin{eqnarray}
 \beta\equiv \frac{{\mathfrak c}}{|{\mathfrak p}|^{1/2}}=\gamma H, \qquad |{\mathcal V}|=|{\mathfrak p}|^{3/2}=a^3,
\end{eqnarray}
with Poisson bracket $\{\beta,{\mathcal V}\}=\frac{\gamma}{2}\mbox{sgn}({\mathcal V})$.

To built the quantum theory in LQC, it is usual to choose as a Hilbert space,
the  quotient space  to the  Besicovitch space of
almost periodic functions by its subspace of null functions. The Besicovitch  space (see for details \cite{b54})
is the closure of  trigonometric polynomials under the
semi-norm (in the ${\mathfrak c}$-representation)
$$||\Psi||^2=\lim_{L\rightarrow \infty}\frac{1}{2L}\int_{-L}^{L} |\Psi({\mathfrak c})|^2 d{\mathfrak c},
$$ where ${\mathfrak c}$ is the connection. And all the  element of this space have the expansion
$$\Psi({\mathfrak c})=\sum_{n\in\Z}\alpha_n |\mu_n\rangle\equiv\sum_{n\in\Z}\alpha_ne^{i\mu_n{\mathfrak c}/2},$$
with $\mu_n\in\R$ and $\alpha_n\in {\mathfrak l}^2$ (the space of square-summable sequences).

\begin{remark}
 It is important to realize that $||.||$ is a semi-norm. For example, all continuous functions with compact support have zero norm.
\end{remark}

In this space on can define the operator $\hat{{\mathfrak p}}$ as follows $\hat{{\mathfrak p}}=-\frac{i\gamma}{3}\frac{d}{d{\mathfrak c}}$.
However the  operator $\hat{\mathfrak c}$ defined by $\hat{\mathfrak c}\Psi({\mathfrak c})={\mathfrak c}\Psi({\mathfrak c})$ is not well
defined in this Hilbert space because for a general quantum state $\Psi({\mathfrak c})=\sum_{n\in\Z}\alpha_ne^{i\mu_n{\mathfrak c}/2}$
one has
$$||\hat{\mathfrak c}\Psi||^2=\lim_{L\rightarrow \infty}\frac{L^2}{3}\sum_{n\in\Z}|\alpha_n|^2=+\infty.$$

Note that, one can also use the $\beta$ representation, where now the semi-norm is
$$||\Psi||^2=\lim_{L\rightarrow \infty}\frac{1}{2L}\int_{-L}^{L} |\Psi({\beta})|^2 d{\beta},$$
and the expansion of quantum states is
\begin{eqnarray}\label{eq1}
\Psi({\beta})=\sum_{n\in\Z}\alpha_n |\nu_n\rangle\equiv\sum_{n\in\Z}\alpha_ne^{i\nu_n{\beta}/2}.
\end{eqnarray}

In this representation the operator $\hat{\mathcal V}$ is given by $-\frac{i\gamma}{2}\frac{d}{d{\beta}}$, but $\hat{\beta}$, for
the same reason as $\hat{\mathfrak c}$,
is not well-defined.

Note that, in the $\beta$ representation, the volume operator, namely $\hat{V}$, is given by $\hat{V}\equiv |\hat{\mathcal V}|$, and act as follows:
$$\hat{V}|\nu\rangle=\frac{\gamma}{4}|\nu||\nu\rangle. $$

\begin{remark}
It is common, in loop quantum cosmology, to use the $\nu$-representation that could be defined as follows:
$$\Psi(\nu)\equiv \frac{1}{4\pi}\int_{\R}\Psi(\beta) e^{-i\nu{\beta}/2}d\beta,$$
and thus, for a general state defined by (\ref{eq1}) one has $\Psi({\nu})=\sum_{n\in\Z}\alpha_n
 \delta(\nu-\nu_n)$.

In this representation, the operator $\hat{\mathcal V}$ acts as follows:
$$\hat{\mathcal V}\Psi(\nu)=\frac{1}{4\pi}\int_{\R}-\frac{i\gamma}{2}\frac{d\Psi(\beta)}{d{\beta}} e^{-i\nu{\beta}/2}d\beta=
\frac{\gamma}{4}\nu\Psi(\nu),$$
where we have used integration by parts.
And thus, $\hat{ V}\Psi(\nu)=
\frac{\gamma}{4}|\nu|\Psi(\nu).$

Finally, the operator $\widehat{e^{i\sigma\beta}}$ acts as follows:
$$\widehat{e^{i\sigma\beta}}\Psi(\nu)=\frac{1}{4\pi}\int_{\R}e^{i\sigma\beta}\Psi(\beta)e^{-i\nu{\beta}/2}d\beta=\Psi(\nu-2\sigma).
$$

\end{remark}

We can also define negative powers of $\hat{V}$ in the following way: First we use the formula
$$ |{\mathcal V}|^{-1/2}=\frac{4i}{\lambda\gamma}e^{i\lambda\beta}\{e^{-i\lambda\beta},|{\mathcal V}|^{1/2}\},$$
where $\lambda$  is a
parameter with dimensions of length, which is determined invoking
the quantum nature of the geometry, that is, identifying its square with the
minimum eigenvalue of the area operator in LQG, which gives as a result $\lambda\equiv
\sqrt{\frac{\sqrt{3}}{4}\gamma}$ (see \cite{s09a}).

Then,  from this formula, we define the operator $\widehat{|{\mathcal V}|^{-1/2}}$, which applied to the eigenstate $|\nu\rangle$ gives

\begin{eqnarray}
\widehat{|{\mathcal V}|^{-1/2}}|\nu\rangle=\frac{2}{\lambda\gamma}\mbox{sgn}({\mathcal V})\left(
\widehat{e^{i\lambda\beta}} [\widehat{e^{-i\lambda\beta}},|\hat{\mathcal V}|^{1/2}]+[\widehat{e^{-i\lambda\beta}},|\hat{\mathcal V}|^{1/2}]
\widehat{e^{i\lambda\beta}}\right)|\nu\rangle \nonumber\\
=\frac{1}{\lambda\sqrt{\gamma}}\left||\nu+2\lambda|^{1/2}-|\nu-2\lambda|^{1/2}|\right|\nu\rangle.
\end{eqnarray}

Note that for large values of $|\nu|$ one has
$$\frac{1}{\lambda\sqrt{\gamma}}\left||\nu+2\lambda|^{1/2}-|\nu-2\lambda|^{1/2}\right|\cong \frac{2}{\sqrt{\gamma}}|\nu|^{-1/2},$$
which coincides with the inverse of the square root of the eigenvalue of the volume operator. And for small values of $\nu$ on has
$$\frac{1}{\lambda\sqrt{\gamma}}\left||\nu+2\lambda|^{1/2}-|\nu-2\lambda|^{1/2}\right|\cong \frac{1}{\sqrt{2\gamma}}\frac{|\nu|}{\lambda^{3/2}},$$
which vanishes at $\nu=0$.

\begin{remark}
 Note that there is an ambiguity in this procedure because one could have started with $|{\mathcal V}|^{-j}$ where $j\in(0,1)$ in
instead of $|{\mathcal V}|^{-1/2}$ (see \cite{as11}).
\end{remark}

Here it is important to stress a key point  in  loop quantum cosmology: The gravitational part of the Hamiltonian contains $\beta$ or ${\mathfrak c}$. In fact, one has
$${\mathcal H}_{grav}=-3H^2a^3=-\frac{3}{\gamma^2}\beta^2|{\mathcal V}|= -\frac{3}{\gamma^2}{\mathfrak c}^2|{\mathfrak p}|^{1/2},$$
and since the operators $\hat{\beta}$ and $\hat{\mathfrak c}$ are not well-defined, in order to built the quantum theory, one needs to
re-define the gravitational part of the Hamiltonian. To be precise, we will work in the $\beta$ representation
and we will consider the holonomies $
h_j(\lambda)\equiv e^{-i\frac{\lambda \beta}{2}\sigma_j}$, where
$\sigma_j$ are the Pauli matrices.
Then, since $\beta^2$ doesn't have a well-defined quantum operator, to construct a consistent  quantum Hamiltonian operator, one needs
one almost periodic function that approaches $\beta^2$ for small values of $\beta$.
This can be done using
the general formulae of loop quantum gravity to
obtain the regularized Hamiltonian
\begin{eqnarray}\label{ham}
&& \hspace*{-5mm}
{\mathcal H}_{LQG}\equiv-\frac{2 |{\mathcal V}|}{\gamma^3 \lambda^3}
\sum_{i,j,k}\varepsilon^{ijk} Tr\left[
h_i(\lambda)h_j(\lambda)h_i^{-1}(\lambda)
\right. \nonumber \\ && \left. \times
 h_j^{-1}(\lambda)h_k(\lambda)\{h_k^{-1}(\lambda),|{\mathcal V}|\}\right]
+\rho |{\mathcal V}|,
\end{eqnarray}
which
 captures the underlying loop quantum dynamics (see \cite{abl03, aps06}). Note that,
this Hamiltonian admits a quantum version because the operators $\hat{h}_j(\lambda)$ are well defined  in this Hilbert space.

\vspace{1cm}

\noindent {\it 4.- Effective dynamics}--- This Section is a based in \cite{h12,bo02a}.
Coming back to the regularized Hamiltonian (\ref{ham}),
an easy calculation gives \cite{he10,dmw09}
\begin{eqnarray}\label{4}
{\mathcal H}_{LQC}=-3|{\mathcal V}|\frac{\sin^2( \lambda \beta)}{\gamma^2\lambda^2}+ \rho|{\mathcal V}|.
\end{eqnarray}

Then, the Hamiltonian constraint is  given by
$\frac{\sin^2( \lambda\beta)}{\gamma^2\lambda^2}=\frac{\rho}{3}$, and the  Hamiltonian equation gives the following identity:
\begin{eqnarray}\label{d}
\dot{{\mathcal V}}=\{{\mathcal V},{\mathcal H}_{LQC}\}=-\frac{\gamma}{2}\frac{\partial{\mathcal H}_{LQC}}{\partial\beta}\mbox{sgn}({\mathcal V})
\Longleftrightarrow \\H= \frac{\sin(2\lambda \beta)}{2\gamma\lambda}\Longleftrightarrow \beta=
\frac{1}{2\lambda}\arcsin(2\lambda\gamma H).
\end{eqnarray}

Writing this last equation as follows $H^2=\frac{\sin^2(\lambda \beta)}{\gamma^2\lambda^2}(1-\sin^2( \lambda \beta))$ and
using the Hamiltonian constraint ${\mathcal H}_{LQC}=0\Longleftrightarrow \frac{\sin^2( \lambda \beta)}{\gamma^2\lambda^2}=\frac{\rho}{3}$
one gets the following modified Friedmann equation in loop quantum cosmology
\begin{eqnarray}\label{6}H^2=\frac{\rho}{3}\left(1-\frac{\rho}{\rho_c}\right)\Longleftrightarrow \frac{H^2}{\rho_c/12}+\frac{(\rho-\frac{\rho_c}{2})^2}{\rho_c^2/4}=1,
\end{eqnarray}
being $\rho_c\equiv \frac{3}{\gamma^2\lambda^2}$.
This equation with the conservation equation $\dot{\rho}=-3H(1+\omega)\rho$ give the dynamics of our universe in loop quantum cosmology.

The integration of these equations shows that if the current energy density of our universe is $\rho(t_0)=\rho_0$,
the dynamics of the universe will be given by
\begin{eqnarray}\label{14}
 \rho(t)=\left(\frac{3}{4}(1+\omega)^2(\bar{t}-t)^2+\frac{1}{\rho_c} \right)^{-1},\qquad H(t)=\frac{1+\omega}{2}(t-\bar{t})\rho(t)
\end{eqnarray}
where $\bar{t}=t_0-\frac{2\sqrt{1-\frac{\rho_0}{\rho_c}}}{\sqrt{3}(1+\omega)\sqrt{\rho_0}}$.
Note that this solution is defined for all time and it  finishes at $(H=0,\rho=0)$, and it
 also shows  that
the universe  expands up to time $t=\bar{t}$,  afterwards  it bounces  and re-collapses forever and ever.

\vspace{0.5cm}

 However this conclusions could be wrong because the Hamiltonian (\ref{4}) is physically unacceptable for the following reason:
It's well-known that the other current cosmological theories are built from two invariant,
the scalar curvature $R=6\left(\dot{H}+2H^2\right)$
and the
Gauss-Bonnet curvature invariant
$G=24H^2\left(\dot{H}+H^2\right)$ (note that the Weyl tensor vanishes for FRW cosmologies). For example,
in Gauss-Bonnet gravity \cite{cenoz06}  the Lagrangian ${\mathcal L}_{MG}=a^3f(R,G)-a^3\rho$ is used,
 and semiclassical gravity, when one takes into account the quantum effects due a massless conformally coupled field (see for instance \cite{hae11}),
is based in the trace anomaly $T_{vac}= \alpha\Box  R-\frac{\beta}{2}G$ (being $\alpha>0$ and $\beta<0$ two re-normalization coefficients).
However applying the Legendre transformation ${\mathcal H}_{LQC}=-\frac{2}{\gamma}\mbox{sgn}({\mathcal V})\dot{\mathcal V}\beta-{\mathcal L}_{LQC}$
one gets,
 in terms of the standard
variables, the following Lagrangian
\begin{eqnarray}\label{l1}
 {\mathcal L}_{LQC}=-\frac{3a^3H}{\gamma\lambda}\arcsin(2\lambda\gamma H)+\frac{3a^3}{2\gamma^2\lambda^2}\left(1-\sqrt{1-4\gamma^2\lambda^2H^2}
\right)-a^3\rho,
\end{eqnarray}
which  is non-covariant, and thus, it contradicts one of the main principles of General Relativity.

As a consequence, since the Hamiltonian (\ref{4}) is obtained from the non-covariant Lagrangian (\ref{l1}),
 one can conclude that this Hamiltonian  is physically unacceptable, and thus,
 the modified Friedmann equation, which was deduced assuming that (\ref{4}) is the Hamiltonian of the system,
 would have been obtained in an incorrect way, and then, all the consequences deduced from it, in particular the big rip singularity avoidance,
 are unjustified and probably incorrect.

Once we have showed the problems with this modified Hamiltonian, it is natural to  wonder, why is used this Hamiltonian  in LQC? The answer seems clear:
 Since the operator $\hat{\beta}^2$ is
not well-defined in  the Besicovitch space
of almost periodic functions one has the replace the standard Hamiltonian by another one built with variables that admit a well-defined
quantum operator in this space. However, another viewpoint is possible. One can  use a different Hilbert space where
$\hat{\beta}^2$ was well-defined. One of these
spaces is ${\mathcal L}^2(SU(2), d\mu_H)$, the space of functions of
isotropic connections which are square integrable with respect to Haar mesure, which
was introduced by Bojowald in \cite{bo02,bo00}. Essentially, this
space, in the ${\mathfrak c}$ representation, is the clause of continuous functions with the norm
$$||\Psi||^2=\int_0^{4\pi}|\Psi({\mathfrak c})|^2\sin^2({\mathfrak c}/2)d{\mathfrak c},$$
and an orthonormal basis is the usual Fourier one, $|n\rangle=\frac{e^{in{\mathfrak c}/2}}{\sqrt{4\pi}\sin({\mathfrak c}/2)}$
with $n\in\Z$.

In this case the operator $\hat{{\mathfrak p}}=-\frac{i\gamma}{3}\left(\frac{d}{d{\mathfrak c}}+\frac{1}{2}\cot({\mathfrak c}/2)\right)$,
which eigenfunctions are $|n\rangle$ and
 has eigenvalues $p_n=\frac{\gamma}{6}n$. It also could be shown that the volume operator has eigenvalues
$$V_{(n-1)/2}= \left(\frac{\gamma}{6}\right)^{3/2}\sqrt{(n-1)n(n+1)}.$$

On the other hand, the eigenvalues of the inverse  volume operator are
difficult to calculate, but they can be approximated by \cite{bo02a}
\begin{eqnarray}
 d_{j,n}\cong\left\{\begin{array}{ccc}
  \frac{12^6}{7^6}V^{-1}_{(n-1)/2}\left(\frac{n}{2j}\right)^{15/2} &\mbox{when}& n\ll 2j\\
    V^{-1}_{(n-1)/2} &\mbox{when}& n\gg 2j,
                    \end{array}\right.
\end{eqnarray}
where $j\in \N/2$ is an ambiguity parameter. (Note that this ambiguity appears because $0$ is an eigenvalue of the volume operator).

Now to obtain the effective Friedmann equation we use Bojowald's ideas (see for instance, \cite{bo02a,bo05}): We
identify the eigenvalues of $\hat{{\mathfrak p}}$ with $a^2$ to
obtain the relation $a^2=\frac{\gamma}{6}n$, then the eigenvalues of the inverse volume operator, as a function of $a$, can be written as follows
\begin{eqnarray}
 d_{j}(a)\cong\left\{\begin{array}{ccc}
  \frac{12^6}{7^6}\left(\frac{3}{\gamma j}\right)^{15/2}a^{12} &\mbox{when}& a^2\ll \gamma j/3\\
    a^{-3} &\mbox{when}& a^2\gg \gamma j/3.
                    \end{array}\right.
\end{eqnarray}

Then, the idea is to  change the Friedmann equation
$$H^2=\rho/3\Longleftrightarrow \frac{\dot{a}^2}{a^2}=\frac{1}{3}\rho_0\left(\frac{a}{a_0}\right)^{-3(1+\omega)},$$
by the effective Friedmann equation
$$\frac{\dot{a}^2}{a^2}=\frac{1}{3}\rho_0\left(\frac{1}{a_0}\right)^{-3(1+\omega)}d_j^{1+\omega}(a),$$
which coincides with the classical Friedmann one for large values of $a$, and thus, the big rip singularity ($\omega<-1$)
is not avoided because
it occurs when $a\rightarrow \infty$.

On the other hand, when  ($\omega>-1$) there is a big bang singularity, which occurs when
$a\rightarrow 0$. In this case, near the singularity, the modified Friedmann equation models the classical one, but with the following EoS,
$P=\omega_{eff}\rho$ where  now $\omega_{eff}=-5-4\omega<-1$ because $\omega>-1$.  This means that  the big bang singularity is avoided, because
this singularity only appears when $\omega_{eff}>-1$ .

Note that all these calculation can be done in an easier way using the $\beta$-representation and
assuming that $\beta$ takes values in a circle. For example, choosing as a Hilbert space
${\mathcal L}^2(-A,A)$, the operator $\hat{\mathcal V}$ is given by $-\frac{i\gamma}{2}\frac{d}{d\beta}$ and  the states
$|n\rangle\equiv \frac{e^{i\pi\beta n/A}}{\sqrt{2A}}$ are orthonormal eigenfunctions with eigenvalues $\frac{\pi\gamma n}{2A}$. The parameter
$A$ could be determined identifying $\frac{\pi\gamma}{2A}$ with $\lambda^3$, which gives
$A=\frac{\pi\gamma}{2\lambda^3}=\frac{4\pi}{\sqrt{\gamma}3^{3/4}}$.
The eigenvalues of the volume and inverse volume operators are  respectively $V_n=\lambda^3|n|$ and
$d_n=\frac{1}{\lambda^2{\gamma}}\left||\frac{4\lambda^3n}{\gamma}+2\lambda|^{1/2}-|\frac{4\lambda^3n}{\gamma}-2\lambda|^{1/2}\right|^2$. Then, performing the
identification $a^3=\lambda^3n$ one gets
\begin{eqnarray}
 d(a)\cong\left\{\begin{array}{ccc}
 \frac{8}{\gamma^3\lambda^3}a^{6} &\mbox{when}& a^3\ll \gamma \lambda/2\\
    a^{-3} &\mbox{when}& a^3\gg \gamma \lambda/2,
                    \end{array}\right.
\end{eqnarray}
and now the modified Friedmann equation, when $\omega>-1$, near the big bang singularity models the classical one with the EoS
$P=\omega_{eff}\rho$ where   $\omega_{eff}=-3-2\omega<-1$, and thus, the big bang singularity is avoided, but not the big rip one.

\begin{remark} Other boundary conditions are also possible in bounded domains, for example, homogeneous Dirichlet boundary conditions (physically it
means that
the Hubble parameter, or the connection, are into an infinite square well, which of course, does not have any physical basis). In this case,
zero is not a eigenvalue of the volume, which allows us to define, without any ambiguity, the eigenvalues of the inverse volume
operator as the inverse of the eigenvalues
of the volume operator.
\end{remark}

The problem in Bojowald's formulation is that one has  assumed that the values of the connection,  or the Hubble parameter, belong in a
circle rather in the real line (recall that in bounded domains a well-defined differential operator needs
boundary conditions, in this case it has been assumed that they were periodic, which means that
the variable takes values in a circle). This assumption has no physical basis and for this reason it was left aside and it emerged the idea that
the
Besicovitch space could be the new  Hilbert space in LQC. However as we have seen throughout the paper, this Hilbert space requires a new
definition of the Hamiltonian operator, and this new Hamiltonian used in LQC doesn't seem to be the correct one, because his classical analogue, after
Lengendre's transformation, gives rise to a non covariant Lagrangian. Moreover, using this Hamiltonian one obtains a modified Friedmann equation
where the Hubble parameter is once again constrained to be bounded, and what is worse, in this modified Friedmann equation
it doesn't appear any higher-curvature term, like  derivatives of the Hubble parameter, which are essential in modified or semiclassical
cosmology.

\vspace{0.5 cm}

\noindent{\it 5.- Phantom cosmology}---

In this Section we will study a toy model where the energy density of the universe is dominated by a phantom scalar field under the action of an
exponential potential. This model allows us to compare
the classical solutions, which give rise to a big rip singularity \cite{dks06,pp09,hae12}, with the solutions obtained in loop quantum cosmology and
the ones obtained
in the framework of the Wheeler-DeWitt equation.

To simplify the calculations we choose the units
$\hbar=c=1$ and $M_p=1/\sqrt{6}$, and the exponential potential $V(\phi)=9e^{-6\phi}$. Then,
the energy density and pressure for this phantom scalar field are
$\rho=-\frac{1}{2}\dot{\phi}^2+9e^{-6\phi}$ and
$P=-\frac{1}{2}\dot{\phi}^2-9e^{-6\phi}$. The
Friedmann and conservation equations read:
\begin{eqnarray}\label{22}\left\{\begin{array}{ccc}
H^2&=& 2\left(-\frac{1}{2}\dot{\phi}^2+9e^{-6\phi}\right),\\
&&\\
0&=&-\ddot{\phi}-3H\dot{\phi} -54e^{-6\phi}.\end{array}\right.
\end{eqnarray}

This system has a global late time attractor solution (all the solutions approach to this one at late times \cite{hae12}) given by
\begin{eqnarray}\label{23}
\phi(t)=\frac{1}{3}\ln\left[1-3H_0(t-t_0) \right];\quad a(t)=\left[1-3H_0(t-t_0) \right]^{-1/3},
\end{eqnarray}
where $H_0$ is the value of the Hubble parameter at time $t_0$.
Introducing $\alpha=\ln(a)$, this solution has the simple form $\phi(\alpha)=-\alpha$.

Note that this solution gives rise to a big rip singularity because when $t\rightarrow t_0+\frac{1}{3H_0}$ the scalar factor, the energy
density and the pressure diverge.

To study the modifications that produce quantum effects,
first at all, we will deduce and study the Wheeler-DeWitt equation for this model. Starting from  the Lagrangian of the system in the
variables $(\alpha,\phi)$
\begin{eqnarray}\label{24}
{\mathcal L}=e^{3\alpha}\left(-\frac{1}{2}(\dot{\alpha}^2+\dot{\phi}^2)-9e^{-6\phi} \right),\end{eqnarray}
and  performing the following change of variable
\begin{eqnarray}\label{25}u(\alpha,\phi)=e^y(\cos x-\sin x);\quad v(\alpha,\phi)=e^y(\cos x+\sin x),
\end{eqnarray}
where $x=3(\alpha+\phi)$ and $y=3(\alpha-\phi)$,
after an straightforward calculation one obtains the following Hamiltonian:
\begin{eqnarray}\label{26}{\mathcal H}=A(u,v)(p_u^2+p_v^2-1),\end{eqnarray}
where $p_u$ and $p_v$ are the corresponding conjugate momenta and $$A(u,v)=-9(u^2+v^2)^{3/4}e^{-\arctan \left(\frac{v-u}{v+u}\right)}.$$

The Hamilton equations are
$$\dot{u}=2A(u,v)p_u, \quad \dot{v}=2A(u,v)p_v, \quad \dot{p_u}=\dot{p_v}=0 \Longrightarrow
 \frac{du}{dv}=\frac{p_u}{p_v}=constant.$$

Then,
 using the Hamiltonian constraint $p_u^2+p_v^2-1=0$ one finally obtains the solutions
\begin{eqnarray}\label{27} u=\frac{-k}{\sqrt{1-k^2}}v+u_0,\end{eqnarray}
where $k$ and $u_0$ are parameters. Note that, the global late time attractor solution correspond to $k=-1/\sqrt{2}$ and $u_0=0$.


On order to obtain the Wheeler-DeWitt equation we
perform the replacement $p_u\rightarrow \partial_u$ and
$p_v\rightarrow \partial_v$ in the constrain $p_u^2+p_v^2-1=0$ to get
\begin{eqnarray}\label{28}\partial_{u^2}^2\Psi+\partial_{v^2}^2\Psi+\Psi=0.\end{eqnarray}

In \cite{dks06} it was found that the  solutions of the Wheeler-DeWitt equation are regular in all the plane $(u,v)$, and  that a wave-packet solution,
peaked  around the classical solution,
spreads as it approaches to the region of big rip singularity, indicating the break down of the semiclassical approximation and the absence of a
classical time. Then, the authors of \cite{dks06} interprets this result as a big rip singularity
avoidance. This reasoning doesn't seem conclusive (see for example \cite{pp09,bl06} where the authors
 claim that Wheeler-DeWitt equation displays a big rip singularity like in the classical model) because, for example,
 wave-packets peaked around a classical trajectory describing
a free quantum particle
spreads due to the increasing of the uncertainity as time passes, but the average of the position and momentum follows the classical trajectories.
This is the criterium followed in loop quantum cosmology when singularities are studied: One has to calculate the quantum evolution
of some expectation values
and compare its values with their classical analogues (see for details \cite{aps06}).

In our case the Wheeler-DeWitt equation has to be transformed in a Schr\"odiger one, in order to guarantee the conservation of the norm.
The way to do this is to take
 the square root to obtain:
\begin{eqnarray}\label{29}
i\partial_v\Psi=\sqrt{1+\partial_{u^2}^2}\Psi,
\end{eqnarray}
where the square root of the self-adjoint operator $1+\partial_{u^2}^2$ has to be understood as the Taylor's expansion of the function
$\sqrt{1+x}=1+\frac{1}{2}x+\dots$ evaluated at $x= \partial_{u^2}^2$.

\begin{remark}
Note that, taking the square root we have chose as a external time, the variable $v$. Then, the time $v$ could be used to define both, the classical  (\ref{27})
and the quantum (\ref{29}) dynamics. Moreover, with this choice one does not need the semiclassical approximation to recover a classical time, because one already has
the time $v$.
\end{remark}

 Since the eigenfunction of this operator are $e^{iku}$ and its eigenvalues
are $\sqrt{1-k^2}$ with $|k|\leq 1$, the solution of equation (\ref{29}) is
\begin{eqnarray}\label{30}
\Psi(u,v)=\frac{1}{\sqrt{2\pi}}\int_{-1}^{1}e^{iS(k;u,v)}A(k)dk,
\end{eqnarray}
where $S(k;u,v)=k(u-u_0)-\sqrt{1-k^2}v$. If we choose as amplitude a Gaussian centered around a point $k=\bar{k}$, with width $\sigma\ll 1$
$$A(k)=\frac{1}{\sqrt{\pi}\sigma}e^{-\frac{(k-\bar{k})^2}{2\sigma^2}},$$ we will be able to approximate (\ref{30}) by
$\Psi(u,v)\cong\frac{1}{\sqrt{2\pi}}\int_{\R}e^{iS(k;u,v)}A(k)dk$. In that case, expanding $S(k;u,v)$ around $\bar{k}$ and neglecting terms of the
order $(k-\bar{k})^3$ one obtains:

\begin{eqnarray}\label{31}
 |\Psi(u,v)|^2=\frac{1}{\pi\sigma^2\sqrt{\frac{1}{\sigma^4}+\frac{v^2}{(1-\bar{k}^2)^{3/2}}}}
\exp\left({\frac{ u+\frac{\bar{k}}{\sqrt{1-\bar{k}^2}}v-u_0}{2\sqrt{\frac{1}{\sigma^4}+\frac{v^2}{(1-\bar{k}^2)^{3/2}}}}}\right),
\end{eqnarray}
which means that $\langle u\rangle=-\frac{\bar{k}}{\sqrt{1-\bar{k}^2}}v+u_0$, that is, $\langle u\rangle$ follows the classical trajectories, and
thus, following this point of view one can conclude that the Wheeler-DeWitt doesn't avoid the big rip singularity.

Once we have show that the big rip survive in the Wheeler-DeWitt theory,  we study
the modified
Friedmann equation in loop quantum cosmology which  in
 the units used in this section reads
\begin{eqnarray}\label{32}
 H^2=2\rho\left(1-\frac{\rho}{\rho_c}\right)\Longleftrightarrow \frac{H^2}{\rho_c/2}+\frac{(\rho-\rho_c/2)^2}{\rho_c/2}=1,
\end{eqnarray}
where now $\rho_c=\frac{1}{2\lambda^2\gamma^2}$.

As we will see from equation (\ref{32}) the energy density belongs in the interval $[0, \rho_c]$ and the Hubble parameter in
$[-\rho_c/2,\rho_c/2]$, so there isn't big rip in the effective formulation of loop quantum cosmology.

Finally, the quantum equation in loop quantum cosmology is very much complicated than equation (\ref{29}), and it is impossible to obtain an analytic
expression like (\ref{31}). But one can reason as follows: If the big rip exists it would be produced in the expanding phase ($H>0$), then we will
only study equation (\ref{32}) in that phase. Taking the derivative with respect the time one obtains
\begin{eqnarray}\label{33}
 \dot{H}=3\dot{\phi}^2\left(1-\frac{2\rho}{\rho_c}\right),
\end{eqnarray}
where we have used the conservation equation  $\dot{\rho}=3H\dot{\phi}^2$. Then since $\rho$ belongs in the interval $[0, \rho_c]$, one has
\begin{eqnarray*}
 |\dot{H}|\leq 3\dot{\phi}^2=\frac{1}{3}\frac{\dot{\psi}^2}{{\psi}^2},
\end{eqnarray*}
where we have introduced the new variable $\psi=e^{3\phi}$.

On the other hand, from the relations $0\leq \rho\leq \rho_c$ one obtains  $\dot{\psi}^2\leq 162$ and
$\frac{1}{\psi^2}\leq \frac{18\rho_c}{162-\dot{\psi}^2}$, which give us the following bound
\begin{eqnarray}\label{34}
 |\dot{H}|\leq \frac{972\rho_c}{162-\dot{\psi}^2}.
\end{eqnarray}

Note that $\rho= \frac{162-\dot{\psi}^2}{18\psi^2}$, which means that $\dot{\psi}$ belongs in the interval $[-\sqrt{162},\sqrt{162}]$.  Then if we
show that $\dot{\psi}^2$ decrease near $\dot{\psi}^2=162$,  $|\dot{H}|$ will remain bounded, that is, the scalar curvature will remain bounded, and thus,  the
the wave function in loop quantum cosmology will exhibit the same behavior as
 the solutions of  the modified Friedmann equation (\ref{32}). As a consequence,
since in the effective formulation the big rip is avoided, it is also avoided in loop quantum cosmology.

To show that $\dot{\psi}^2$ decrease, we write the equation $-\ddot{\phi}-3H\dot{\phi}-54 e^{-6\phi}=0$ in terms of $\psi$, to get
\begin{eqnarray}\label{35}
 \frac{d}{dt}(\dot{\psi})=-\frac{\sqrt{162-\dot{\psi}^2}}{\psi}\left[\sqrt{162-\dot{\psi}^2}+9 \dot{\psi} \right],
\end{eqnarray}
which means that near the points $\dot{\psi}^2=162$ one has
\begin{eqnarray}\label{36}
 \frac{d}{dt}(\dot{\psi})\cong\frac{-9\dot{\psi}\sqrt{162-\dot{\psi}^2}}{\psi}.
\end{eqnarray}

From this equation one can deduce that when $\dot{\psi}$ is near to $\sqrt{162}$ (resp. $-\sqrt{162}$) it decreases (resp. increases), that is
$\dot{\psi}^2$ decreases near $\dot{\psi}^2=162$, and consequently, the scalar curvature remains bounded.

To conclude this Section: We have shown that the Wheeler-DeWitt equation displays the same dynamics as your classical analogue given by the Hamiltonian
(\ref{1}), and consequently the big rip singularity survives in these formulations. On the other hand, loop quantum cosmology displays the
same dynamics as the Hamiltonian (\ref{4}), and thus, one might conclude that the big rip singularity is avoided in loop quantum cosmology.
The difference between these
behaviors comes from the different expressions of  the Hamiltonian (\ref{1}) and (\ref{4}). The first one is physically acceptable because it
comes from a
covariant Lagrangian (it is linear in the curvature ${\mathcal L}=\frac{1}{2}Ra^3-\rho a^3$), but the second one is un-physical because it comes
from the Lagrangian (\ref{l1}) which is non-covariant. This means that the conclusions obtained in loop quantum cosmology
about the big rip singularity avoidance are not acceptable, that is, one cannot conclude that this theory avoids this singularity.

Finally note that the difference between the Wheeler-DeWitt equation and its analogue in loop quantum cosmology near the big rip, comes from the
difference between their classical analogues (the Hamiltonian (\ref{1}) and (\ref{4})), because they are only their quantum versions. More precisely,
the Hamiltonian constrain (\ref{a2}) gives a parabola in the plane $(H,\rho)$ which allow to take unbounded values to the variables $H$ and $\rho$,
and thus, to reach the big rip singularity. On the other hand,  in loop quantum cosmology the Hamiltonian constraint (\ref{6}) gives an ellipse
in the plane $(H,\rho)$,
 which
prevent the big rip singularity. That is, in classical cosmology the universe
is constrained to move along a parabola, and in loop quantum cosmology it moves along an ellipse. Both curves are only similar for small values of
$H$ and $\rho$, but for large values (this means near singularities) they are completely different. This is the main difference between classical  and loop quantum cosmology.

\vspace{0.5 cm}

\noindent{\it 6.- Conclusions}---
Our main conclusion   is that in loop quantum cosmology, contrary to the current belief, the big rip singularity is not replaced by a non-singular bounce.
What really happens is the theory is built from a Hamiltonian that, after a Legendre's transformation, gives a non covariant Lagrangian
which is in contradiction with the principles of  General Relativity. Then, all the conclusions obtained from this Hamiltonian, in particular
the big rip singularity, could be wrong.
Another way to understand that the big rip singularity must survive in loop quantum cosmology is the fact that this theory is based in the discrete
nature of the space, which could influence the dynamics of our universe at small distances (at the order of Planck scale), but never at large distances
where the discrete  space could be substituted by a continuous one, and thus, near the big rip the dynamics of the universe is unaffected by
loop quantum corrections.

\vspace{.5 cm}

\noindent{\bf Acknowledgments.} I would like to thank the comments of the referee, that has been very important in order to write Section V and improve the work.
This investigation has been
supported in part by MICINN (Spain), projects MTM2011-27739-C04-01,
and by AGAUR (Generalitat de Ca\-ta\-lu\-nya), contract 2009SGR-345.

\end{document}